\pdfoutput=1
\documentclass[10pt,twocolumn,twoside]{IEEEtran}
\usepackage{amsmath,graphicx,amssymb}
\usepackage{times}
\usepackage{graphicx} 
\usepackage{pinlabel,subcaption}
\usepackage{url}
\usepackage{color,cite}
\usepackage{algorithmic,algorithm}
\def\x{\underline{x}_j}

\DeclareMathOperator*{\argmax}{arg\,max}

\title{Who Spoke What? A Latent Variable Framework for the Joint Decoding of Multiple Speakers and their Keywords}
\author{Harshavardhan Sundar \and{and} Thippur V. Sreenivas
 \thanks{H. Sundar is with the Language Technologies Institute, Carnegie Mellon University, Pittsburgh, PA 15213. T. V. Sreenivas is with the Department of Electrical Communication Engineering, Indian Institute of Science, Bangalore, India. This work was done in the Indian Institute of Science, Bangalore, India.\newline Emails: suharsha@cs.cmu.edu, tvsree@ece.iisc.ernet.in. 
 }}
\begin{document}
\maketitle
\begin{abstract}
In this paper, we present a latent variable (LV) framework to identify all the speakers and their keywords given a multi-speaker mixture signal. We introduce two separate LVs to denote active speakers and the keywords uttered. The dependency of a spoken keyword on the speaker is modeled through a conditional probability mass function. The distribution of the mixture signal is expressed in terms of the LV mass functions and speaker-specific-keyword models. The proposed framework admits stochastic models, representing the probability density function of the observation vectors given that a particular speaker uttered a specific keyword, as speaker-specific-keyword models. The LV mass functions are estimated in a Maximum Likelihood framework using the Expectation Maximization (EM) algorithm. The active speakers and their keywords are detected as modes of the joint distribution of the two LVs. In mixture signals, containing two speakers uttering the keywords simultaneously, the proposed framework achieves an accuracy of $82\%$ for detecting both the speakers and their respective keywords, using Student's-t mixture models as speaker-specific-keyword models. 
\end{abstract}

\section{Introduction}
\label{sec:intro}

Human conversations, quite often, have multiple people talking at the same time. Automatic processing of such conversations is essential in the context of human-machine interaction (HMI), thus enabling the machine to aid humans better. Consider for example,  a home environment wherein multiple people require the machine to do different things. In such a scenario it becomes important for the machine to understand who spoke what. This requires the machine to be able to identify multiple speakers, recognize speech from multiple speakers and associate the recognized speech streams to the corresponding speaker. 

The problem of speaker recognition has effective solutions \cite{Birnbaum96,Farrell94,Grimaldi08,Nakagawa12,Reynolds95,Wasson75,Xugang07,harsha10}, which are robust to reverberation \cite{Zhao14}, environmental noise \cite{Ming07}, large population \cite{Yakun13} etc. However, recognizing a target speaker in multi-speaker scenario is a harder problem with fewer solutions \cite{Hershey201045}. An even more challenging problem is that of identifying multiple speakers in a multi-speaker scenario \cite{harsha12}.

Although there are robust algorithms and systems for speech recognition in a single speaker scenario \cite{Hinton12,Sainath13,Hannun14,Hamid14,Li14,Stern12,Chiu12,Raj05,Seltzer06,Singh02}, recognizing speech in a multi-speaker scenario, is still a challenge. In order to compare different approaches to recognize speech from a target speaker in a multi-speaker scenario, Cooke et. al. constituted the ``Monaural Speech separation and recognition challenge" \cite{Cooke10} , wherein, the sentences from several speakers were recorded with restricted grammar. The task was to recognize the letter and digit spoken by a speaker who spoke the word ``white", in a multi-speaker mixture signal, wherein the target speaker is masked by another speaker uttering a similar sentence, without the word ``white". Although unrealistic with a restricted sentence grammar, the task was still challenging with only a few approaches able to surpass the human performance \cite{Hershey201045,weng14} (albeit in a sub category). In \cite{Barker2010,sadjadi2014,heck2004}, the authors, address the problem of identifying the target speaker and his keyword, in multi-speaker mixture signal. As noted by the authors, the performance is generally poor at a signal to interference ratio (SIR) of $0$ dB. 

Thus, in general, the task of recognizing speech from multiple speakers is a complex problem. Added to this, in order to address the problem of who spoke what, it is required to associate the recognized speech segments to their respective orators, which is an even more complicated task. 

As noted in \cite{Shneiderman80,Baber91}, the use of restricted vocabulary of keywords may be more suitable for a task driven application, like HMI in a home environment, where the primary concern is to control the machine, for the efficient completion of a task. Therefore, the problem now is formulated to detect, which of the known speakers uttered which of the known keywords. The goal is to identify all the speakers and their keywords rather than keyword from a single target speaker. 

In this paper, we propose to address the above task in a Latent Variable (LV) framework. We associate one LV to denote the active speaker and another LV to denote the keyword uttered and relate the two LVs through a conditional dependency. The probability mass function (p.m.f.) and the conditional p.m.f of the LVs are estimated using speaker specific keyword models. In order to evaluate the proposed framework in relevance to our goal of HMI in a home environment, we have created our own database \footnote{The entire database is available for research purpose only and available on contacting the corresponding author. }. With Student's-t Mixture Models (tMMs) as speaker specific keyword models, the proposed approach is able to detect at least one speaker-keyword pair, in mixture signal with two speakers, with an accuracy of $99\%$ and both speaker-keyword pairs, with an accuracy of $82\%$. 
 The contributions of this paper are: (i) Formulation of the problem of identifying who spoke what in a LV framework (Sec. \ref{lvfs}); (ii) Casting the problem of LV density estimation in a maximum likelihood (ML) framework and solving the same using an Expectation-Maximization (EM) algorithm (Sec. \ref{lvfs}); (iii) Experimental evaluation of the proposed approach on a newly collected database meant for HMI in a home environment (Sec. \ref{perf_eval}).

\section{Proposed Latent Variable (LV) Formulation}
\label{lvfs}
Let $S_k$ denote the $k^{th}$ speaker in the set $\mathcal{S}=\{S_1, S_2, \ldots, S_M\}$ of $M$ known speakers. The mixture signal -$x[n]$ contains speech from a subset of speakers from the set $S$. In a given signal $x[n]$, a speaker is said to be active if he/she speaks and passive otherwise. We assume that the speakers only utter keywords from the vocabulary $V=\{V_1, V_2,\ldots, V_N\}$. Let $X$ denote the features estimated from $x[n]$; $X=\left[\underline{x}_1, \underline{x}_2, \ldots, \underline{x}_j, \ldots, \underline{x}_T\right]; ~ \x \in \mathcal{R}^D;~X\in \mathcal{R}^{D \times T}$.

We introduce two Boolean LVs, one to denote the active speakers ($\underline{U}_j \in \mathcal{B}^{M\times 1}$), and the other to denote the keywords uttered ($\underline{W}_j\in \mathcal{B}^{N\times 1}$), for each feature vector $\underline{x}_j$. $U_j(k)=1$ iff the $k^{th}$ speaker $S_k$ is active in the $j^{th}$ frame. $W_j(l)=1$ iff the $l^{th}$ keyword has been uttered in the $j^{th}$ frame. Modeling the conditional p.d.f. of $\underline{x}_j$ given multiple active speakers as a sum of the conditional p.d.f. of $\underline{x}_j$ given individual speakers, the p.d.f. of $\underline{x}_j$, is obtained as a marginal of the joint p.d.f. of $\underline{x}_j, \underline{U}_j,$ and $\underline{W}_j$, i.e.,
 \begin{align}
\Pr\left(\underline{x}_j\right) &= \sum\limits_{k=1}^{M}\sum\limits_{l=1}^{N}\Pr\left(U_j(k)=1\right)\Pr\left(W_j(l)=1 \mid U_j(k)=1\right)\nonumber \\
&\Pr\left(\underline{x}_j \mid U_j(k)=1,W_j(l)=1\right)\label{data_model_DLV}
\end{align}
where, $\Pr\left(U_j(k)=1\right)$ denotes the probability of the $k^{th}$ speaker being active in the $j^{th}$ frame, $\Pr\left(W_j(l)=1 \mid U_j(k)=1\right)$ represents the conditional probability of the $l^{th}$ keyword being uttered by the $k^{th}$ speaker in the $j^{th}$ frame, and $\Pr\left(\underline{x}_j \mid U_j(k)=1,W_j(l)=1\right)$ denotes the probability of $\underline{x}_j$ given that in the $j^{th}$ frame the $k^{th}$ speaker uttered the $l^{th}$ keyword. The two LVs are related through the use of conditional p.d.f $\Pr\left(W_j(l)=1 \mid U_j(k)=1\right)$, as our goal is to estimate which speaker spoke which keyword. If the LVs are assumed to be independent, then we will be able to decode the set of active speakers and the set of keywords uttered, but the association of the keyword to the corresponding active speaker becomes a combinatorial problem.

 We use parametric speaker-specific-keyword models to compute $\Pr\left(\underline{x}_j \mid U_j(k)=1,W_j(l)=1\right)$. Let the parameters of the model for the $k^{th}$ speaker uttering the $l^{th}$ keyword be denoted by $\lambda_{kl}$ i.e. $\Pr\left(\underline{x}_j \mid U_j(k)=1,W_j(l)=1\right) \triangleq \Pr\left(\underline{x}_j;\lambda_{kl}\right)$. 
Let $\mathbf{\lambda}$ denote the collection of all speaker specific keyword model parameters,  $\mathbf{\lambda}=\{\lambda_{kl}; 1\leq k \leq M, 1\leq l \leq N\}$. Assuming that the distribution of a particular speaker and the distribution of a speaker uttering a particular keyword are time homogeneous, we represent $\Pr\left(U_j(k)=1\right)$ by $\beta_{k}$ and $\Pr\left(W_j(l)=1 \mid U_j(k)=1\right)$ as $\delta_{kl}$.
Further, assuming that in the given utterance $x[n]$, at least one of the $M$ speakers utters one of the $N$ keywords \footnote{Out of vocabulary (OOV) keywords can be handled using a garbage model for each speaker. Effectively, this increases the vocabulary size to $N+1$. However, the choice of suitable garbage model to effectively detect the OOV words is beyond the scope of this paper.}, we have:

\begin{equation}
\Pr\left(\x;{\mathbf{\lambda}}\right)=\sum\limits_{k=1}^{M}\beta_k \sum\limits_{l=1}^{N}\delta_{kl} ~\Pr\left(\x;\lambda_{kl}\right)
\end{equation}
with$\sum\limits_{k=1}^{M}\beta_k =1;$ and $\sum\limits_{l=1}^N \delta_{kl}=1;~\forall 1 \leq k \leq M$.
Let ${\mathbf{\delta}}=[\delta_{kl}]$, $\underline{\beta}=\left[\beta_1,\ldots,\beta_M\right]$. We propose to solve for the parameters ${\mathbf{\delta}}$ and $\underline{\beta}$, in an ML framework using the EM algorithm \cite{DLR}, keeping the parameters $\lambda$ fixed. 

\subsection{EM algorithm for estimation of $\underline{\beta}$ and  ${\mathbf{\delta}}$}

Let the posterior probabilities at the $m^{th}$ EM iteration be defined as: $\eta_{jk}^{(m)} \triangleq \Pr\left(U_j(k)=1 \mid \x;\lambda_{kl}\right)$,\newline
$\zeta_{jkl}^{(m)} \triangleq  \Pr\left(U_j(k)=1, W_j(l)=1 \mid \x ; \lambda_{kl}\right)$. The Q-function, which is the conditional expectation of the log-likelihood of the overall data w.r.t. the LVs given the data, is derived as:

\begin{align}
Q\left(\psi, \psi^{(m)};{\mathbf{\lambda}}\right)&=\sum\limits_{j=1}^{T}\sum\limits_{k=1}^{M}\sum\limits_{l=1}^{N}\eta_{jk}^{(m)}\ln \left(\beta_k\right) + \zeta_{jkl}^{(m)} \ln \left(\delta_{kl}\right) \nonumber \\
&+ \ln \left(\Pr\left(\x;\lambda_{kl}\right)\right)
\end{align}
where $\psi \triangleq \{{\mathbf{\delta}}, \underline{\beta}\}$ is the collection of parameters to be estimated and $\psi^{(m)}\triangleq \{{\mathbf{\delta}}^{(m)}, \underline{\beta}^{(m)}\}$, is the collection of the parameters at the $m^{th}$ EM iteration. The EM update equations for $\eta_{jk}^{(m)}$ and $\gamma_{jkl}^{(m)}$ are given as:

\begin{align}
\eta_{jk}^{(m)} &= \frac{\beta_k^{(m)}\sum\limits_{l=1}^{N}\delta_{kl}^{(m)} \Pr\left(\x;\lambda_{kl}\right)}{\sum\limits_{k_1=1}^{M}\beta_{k_1}^{(m)}\sum\limits_{l=1}^{N}\delta_{k_1l}^{(m)} \Pr\left(\x;\lambda_{k_1l}\right)}\\
\zeta_{jkl}^{(m)} &= \frac{\beta_k^{(m)}\delta_{kl}^{(m)} \Pr\left(\x;\lambda_{kl}\right)}{\sum\limits_{k_1=1}^{M}\beta_{k_1}^{(m)}\sum\limits_{l_1=1}^{N}\delta_{k_1l_1}^{(m)} \Pr\left(\x;\lambda_{k_1l_1}\right)}
\end{align}
The parameters ${\mathbf{\delta}}^{(m)}$ and $\underline{\beta}^{(m)}$ are updated as:

\begin{align}
\beta_k^{(m+1)}&=\argmax_{\beta_k}\left[Q\left(\psi,\psi^{(m)};{\mathbf{\lambda}}\right)\right];~ \text{s. t.} \sum\limits_{k=1}^{M}\beta_k=1 \nonumber\\
 &=\frac{1}{T}\sum\limits_{j=1}^{T}\eta_{jk}^{(m)}\\ 
\delta_{kl}^{(m+1)}&=\argmax_{\delta_{kl}} \left[Q\left(\psi,\psi^{(m)};{\mathbf{\lambda}}\right)\right];~ \text{s. t.} \sum\limits_{l=1}^{N}\delta_{kl}=1\nonumber\\
&=\frac{\sum\limits_{j=1}^{T}\zeta_{jkl}^{(m)}}{\sum\limits_{j=1}^{T} \sum\limits_{l'=1}^{N}\zeta_{jkl'}^{(m)}}\label{Eq_beta_expr_main}
\end{align}
After convergence of the EM algorithm let the values of $\underline{\beta}$ and ${\mathbf{\delta}}$ be denoted as $\underline{\beta}^*=\left[\beta_1^*,\ldots,\beta_M^*\right]$ and ${\mathbf{\delta}}^*=\{\delta_{kl}^*;~ 1\leq k \leq M; ~ 1\leq l \;\leq N\}$.
Since $\delta_{kl}^*$ is a conditional p.d.f. of the $l^{th}$ word given the $k^{th}$ speaker, we obtain the joint probability matrix ($JPM$), to detect the speaker-keyword pairs, as: $JPM\left(k,l\right)=\beta_k^*.\delta_{kl}^*$. We first pick the active speakers using $\underline{\beta}^*$, and then pick only one peak in the row corresponding to active speakers in $JPM\left(k,l\right)$. From the physics of the problem, although multiple speakers can utter the same keyword, a single speaker cannot utter multiple keywords at the same time. Thus, we pick only one peak in active speaker rows of $JPM\left(k,l\right)$. This is also the reason for choosing the dependency of $W_j(l)$ on $U_j(k)$ in   Eq.(\ref{data_model_DLV}), and not the other way round.

We refer to this proposed framework as latent variable based detection of speakers and keywords or LVDSK in short. 
\subsubsection{Using Prior Knowledge}
\label{sec:prior}
For the estimation of $\beta_k^{(m+1)}$ and $\delta_{kl}^{(m+1)}$, the EM algorithm requires initial estimates $\beta_k^{(0)}$ and $\delta_{kl}^{(0)}$. With no prior knowledge, a flat initialization is used with: $\beta_k^{(0)}=\frac{1}{M}; ~\forall~ 1 \leq k \leq M$, 
$\delta_{kl}^{(0)} = \frac{1}{N}; ~ \forall~ 1 \leq k \leq M; 1 \leq l \leq N$.
However, LVDSK can be effectively used for incorporating any prior knowledge with regards to either the active speakers, or the keywords uttered, effectively, to estimate the unknown better. Let $M^*$ denote the number of active speakers. If the active speakers are known a priori, then $\beta_k^{(0)}$ is set to be $\frac{1}{M^*}$ for active speaker indices and $0$ for passive speaker indices. $\delta_{kl}^{(0)}$ is set to have equal mass ($=\frac{1}{N}$) for all keywords of active speakers and $0$ for all keywords of passive speakers. 
If the keywords uttered are known a priori, then a flat initialization is used for $\beta_k^{(0)}$, and $\delta_{kl}^{(0)}$ is set to have $\frac{1}{M^*}$ for the keywords uttered, for all speakers and 
$0$ for other keywords of all speakers.
In Sec. \ref{sec:perf:res:prior}, we show that, with such prior knowledge LVDSK performs better in estimating the remaining unknown quantities better.
\section{Performance Evaluation}
\label{perf_eval}
\subsection{Database, Features and Performance Measures}
\label{sec:perf:data}
\begin{figure}
 \begin{center}
 \includegraphics[width=2in]{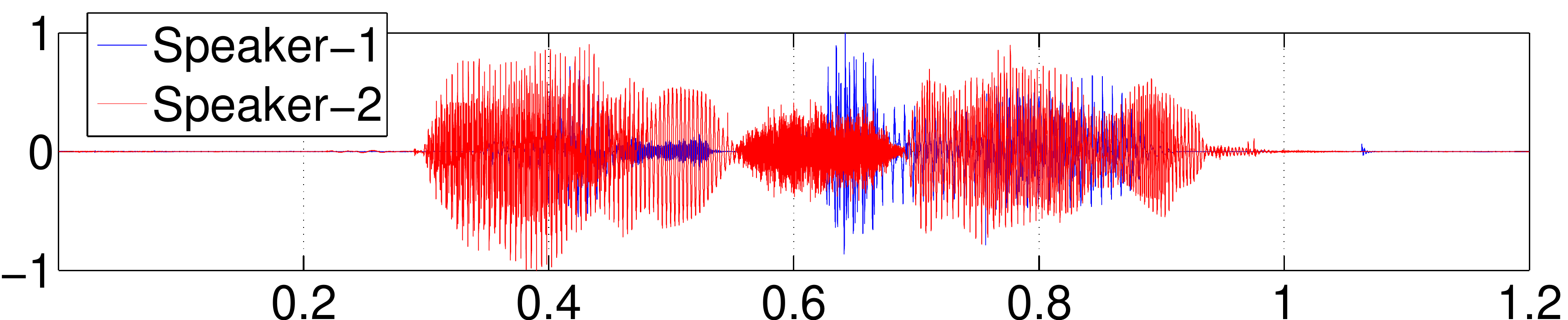}
 \caption{({\it Color Online}): An mixture signal with 2 speakers. } 
 \label{fig_eg_mix_sig}
 \end{center}
\end{figure}

\begin{table}
\begin{center}
\caption{Keywords}
\label{tab_kp}
\begin{tabular}{|c|c||c|c|}
\hline
 Index & Key-Phrase & Index & Key-Phrase \\ \hline 
 1. & Answer & 6.& Music  \\ \hline 
 2. & Disconnect  & 7. & Number  \\ \hline 
 3. & Emergency & 8. &  Outside  \\ \hline
 4. & Hello &  9. & Television   \\ \hline 
 5. & Inside & 10. & Volume   \\ \hline  
\end{tabular}
\end{center}
\end{table}
We have created our own database with the context of HMI in a home environment. Six male speakers, and 4 female speakers are made to utter each of the ten keywords shown in Table. \ref{tab_kp}, with 10 repetitions each. The keywords are chosen carefully to ensure that the vocabulary contains long distinct words (``Emergency", ``Disconnect", ``Television", ``Volume"), smaller distinct words (``Hello", ``Music"), moderately distinct words (``Answer", ``Number" ) and confusable words (``Inside" and ``Outside"). The utterances are recorded using an omni-directional microphone Audio-Technica AT8004L  (For specifications c.f. \cite{mic}), which has a flat frequency response between $80$ Hz to $16$ kHz. Microphone signals are sampled at 16kHz with 16 bits per sample in an anechoic chamber. If the utterances of the two speakers are well separated in time, then, even with very simple keyword spotting techniques, a very accurate detection system can be built. However the problem becomes more difficult when the utterances overlap in time when one keyword masks the other keyword. In order to study this harder problem, albeit unrealistic, the test data is generated by adding the utterances from two different speakers with an overlap of more than $90\%$. In the considered context, since there is no notion of a target and a masker, the measure of SIR is not relevant to characterize the mixture signals. Therefore, we employ a related measure - Relative Power Ratio (RPR), defined as the ratio of the power of each speaker in a mixture signal. All the mixture signals generated, have an RPR close to $0$ dB. An example mixture signal with speech from the two speakers are shown in Fig. \ref{fig_eg_mix_sig}. An example set of test data is available at \url{https://sites.google.com/site/harshas123/downloads}
The performance of the proposed framework is studied under two categories: Multiple speakers uttering different keywords (MSpDKW) and Multiple speakers uttering the same keywords (MSpSKW). With 10 speakers and 10 keywords, there are $\binom{10}{2} \times \binom{10}{2} = 2025$ possible combinations of different speakers uttering different keywords and $\binom{10}{2} \times 10 = 450$ possible combinations of different speakers uttering the same keyword. We therefore generate $2025$ mixture signals for the MSpDKW task and $450$ signals for the MSpSKW task. 
Thirty eight-dimensional Mel Frequency Cepstral Coefficients (MFCCs) along with Delta and Acceleration coefficients, obtained with a frame length of $20$ ms and frame shift of $10$ ms are used as features (omitting the energy component of MFCC). 

LVDSK involves detection of speakers, keywords and the speaker-keyword pairs. To assess the performance of LVDSK, we compute the average percentage recognition of one or both of, each of the three entities - speakers, keywords and speaker-keyword pairs, in a leave one out cross validation setting.
\subsection{LVDSK with different speaker specific keyword models}
\label{sec:perf:gmmvstmm}

Although any stochastic model can be used, we explore the use of two different speaker-specific-keyword models : Gaussian Mixture Models (GMMs) and Student's-t Mixture Models (tMMs). The parameters $\lambda_{kl}$ for both GMMs and tMMs are obtained using the clean speech utterances of the $l^{th}$ keyword by the $k^{th}$ speaker ($S_k$) in an ML framework using an EM algorithm (For GMMs c.f. \cite{bishop2006}. For tMMs c.f. \cite{Peel2000}). For both GMMs and tMMs, $8$ mixture components are used. Table \ref{tab_MM} tabulates the performance of GMMs and tMMs as speaker-specific-keyword models in the LVDSK framework. 
\begin{table*}
\begin{center}
\caption{Overall $\%$ recognition accuracy of LVDSK with GMMs and tMMs as speaker-specific-keyword models.}
\label{tab_MM}
\begin{tabular}{|c|c|c|c|c|c|c|}
\hline
 & At least 1 speaker &  Both speakers &  At least 1 phrase&  Both phrases &  At least 1 speaker-phrase &  Both speaker-phrases\\ 
 & detected correctly & detected correctly & detected correctly & detected correctly & detected correctly & detected correctly \\ \hline
LVDSK-tMM & 99.99 & 91.80 & 99.80 & 83.56 & 99.64 & 81.66 \\ \hline
LVDSK-GMM & 99.40 & 70 & 97.70 & 61 & 96.80 & 57.10 \\ \hline
\end{tabular}
\end{center}
\end{table*}
\begin{table*}
\begin{center}
\caption{Average $\%$ recognition accuracy of LVDSK in different tasks- MSpDKW and MSpSKW with flat initialization.}
\label{tab_tasks}
\begin{tabular}{|c|c|c|c|c|c|c|}
\hline
 Task& At least 1 speaker &  Both speakers &  At least 1 phrase&  Both phrases &  At least 1 speaker-phrase &  Both speaker-phrases\\ 
 (No. of Utterances) & detected correctly & detected correctly & detected correctly & detected correctly & detected correctly & detected correctly \\ \hline
MSpDKW (2025) & 99.99 & 91.33 & 99.80 & 81.80 & 99.65 & 80.50 \\ \hline
MSpSKW (450) & 99.97 & 93.86 & 99.71 & 91.50 & 99.60 & 87.13 \\ \hline
Overall (2475) & 99.99 & 91.80 & 99.80 & 83.56 & 99.64 & 81.66 \\ \hline
\end{tabular}
\end{center}
\end{table*}
\begin{table*}
\begin{center}
\caption{Overall $\%$ recognition accuracy of LVDSK with prior knowledge.}
\label{tab_prior}
\begin{tabular}{|c|c|c|c|c|c|c|}
\hline
 & At least 1 speaker &  Both speakers &  At least 1 phrase&  Both phrases &  At least 1 speaker-phrase &  Both speaker-phrases\\ 
 & detected correctly & detected correctly & detected correctly & detected correctly & detected correctly & detected correctly \\ \hline
LVDSK-Flat & 99.99 & 91.80 & 99.80 & 83.56 & 99.64 & 81.66 \\ \hline
Oracle-SPID & 100 & 100   & 99.81 & 85.22 & 99.74 & 85.22 \\ \hline
Oracle KWID& 100 & 92.3 & 100 & 100 & 100 & 92.3 \\ \hline
\end{tabular}
\end{center}
\end{table*}
Although GMMs and tMMs have nearly comparable performance in detecting at least one entity correctly, tMMs outperform GMMs in detecting both the entities be it speakers, keywords or speaker-keyword pairs. From henceforth, we use tMMs as speaker-specific-keyword models, in all the experiments to follow.


\subsection{Results}
\label{sec:perf:res}
\begin{figure}
\begin{subfigure}[l]{0.23\textwidth}
\labellist
\small\hair 2pt
\pinlabel \rotatebox{-35}{Speaker} [c] at 63 230
\pinlabel \rotatebox{-35}{Index ($k$)} [c] at 33 180
\pinlabel \rotatebox{21}{Word Index ($l$)} [c] at 510 200
\endlabellist
\includegraphics[width=1.2in]{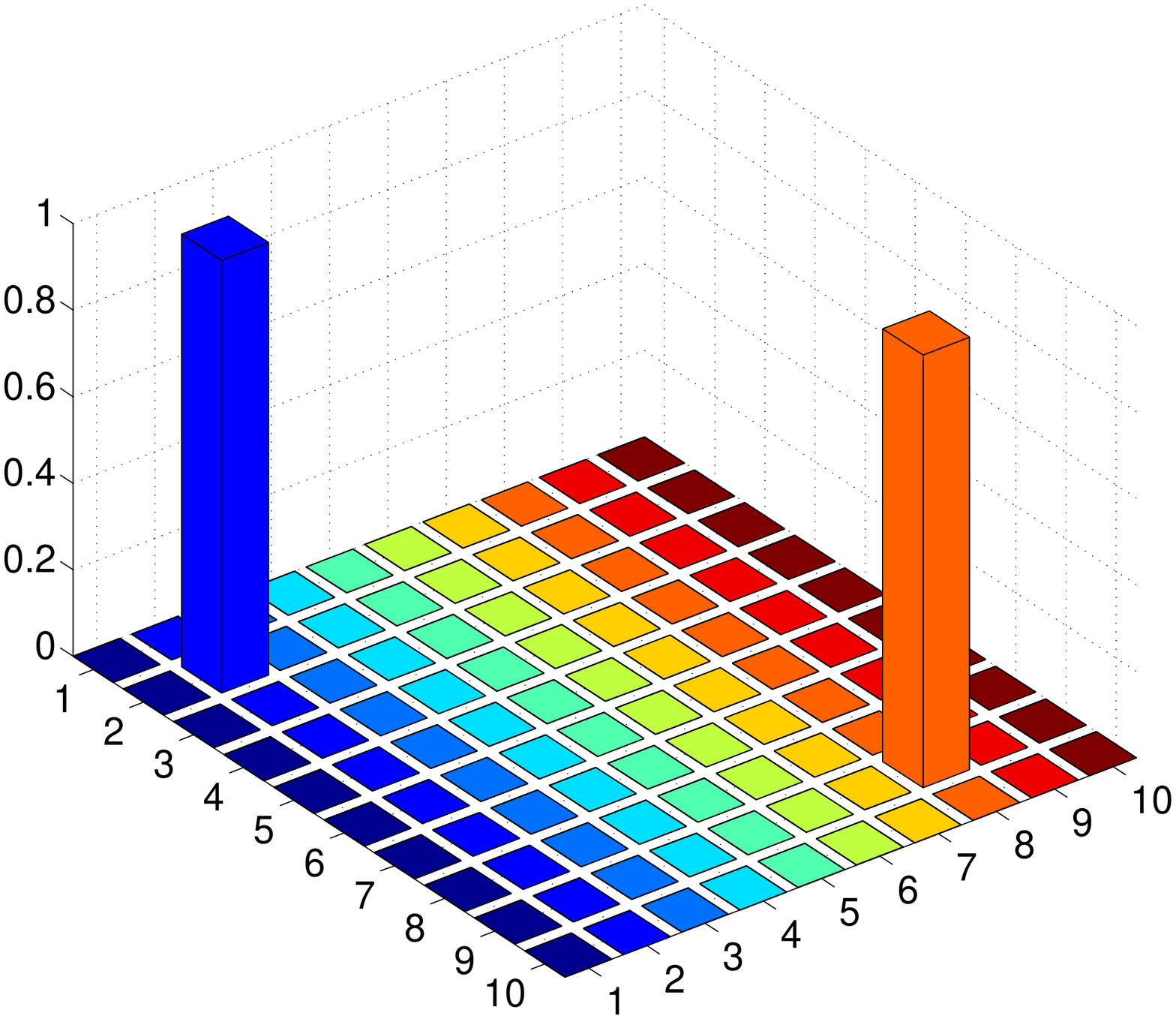}
\vspace*{2mm}
\caption{Ground Truth}
\label{anec_gnd_trth}
\end{subfigure}
\begin{subfigure}[l]{0.23\textwidth}
\labellist
\small\hair 2pt
\pinlabel \rotatebox{-35}{Speaker} [c] at 63 230
\pinlabel \rotatebox{-35}{Index ($k$)} [c] at 33 180
\pinlabel \rotatebox{21}{Word Index ($l$)} [c] at 510 200
\endlabellist
\includegraphics[width=1.2in]{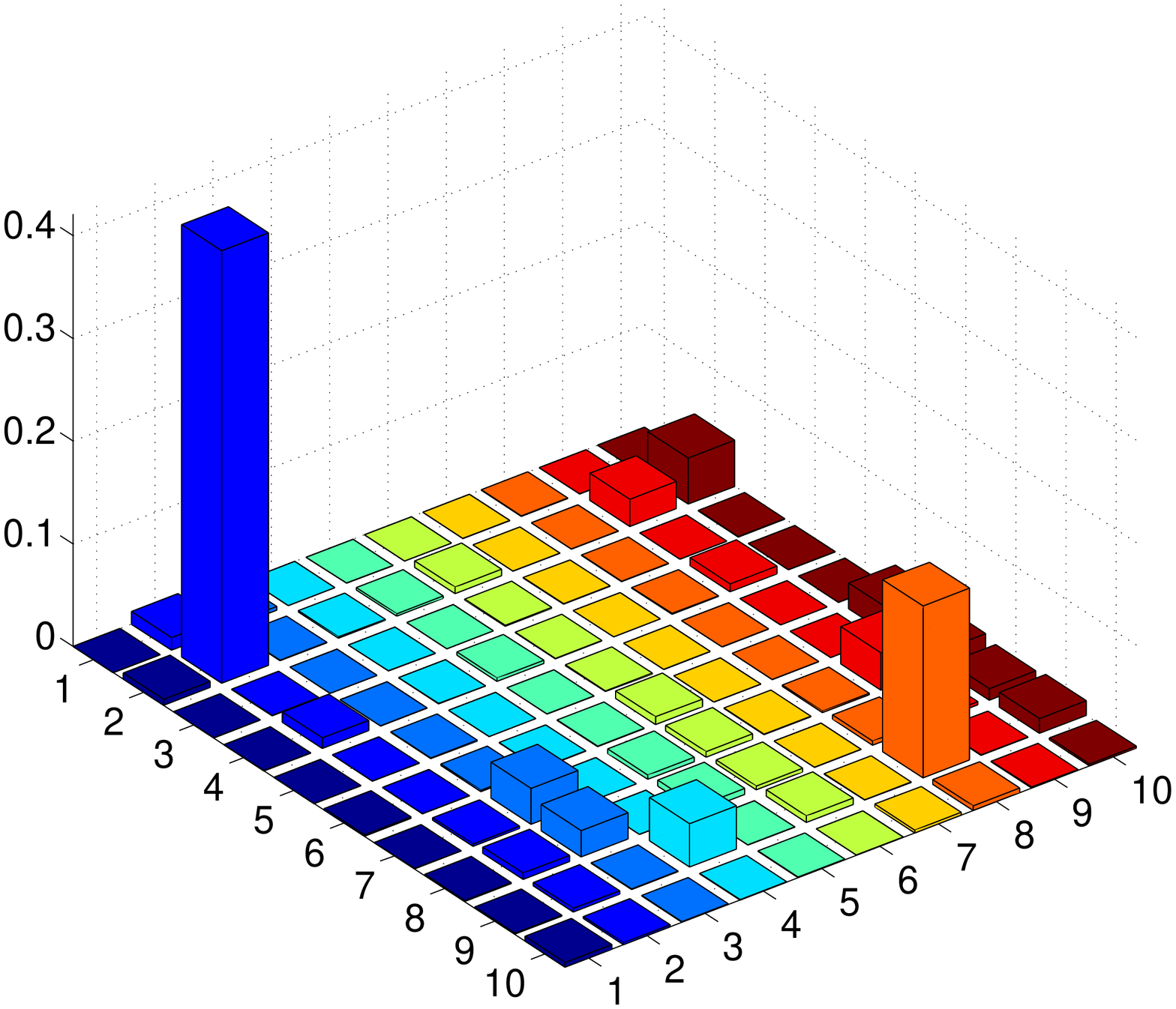}
\vspace*{2mm}
\caption{$JPM\left(k,l\right)$}
\label{anec_delta}
\end{subfigure}
\begin{subfigure}[c]{0.48\textwidth}
\begin{center}
\labellist
\small\hair 2pt
\pinlabel {Speaker Index ($k$)} [c] at 295 490
\pinlabel $\beta_k$ [r] at -190 600
\endlabellist
\includegraphics[width=2in]{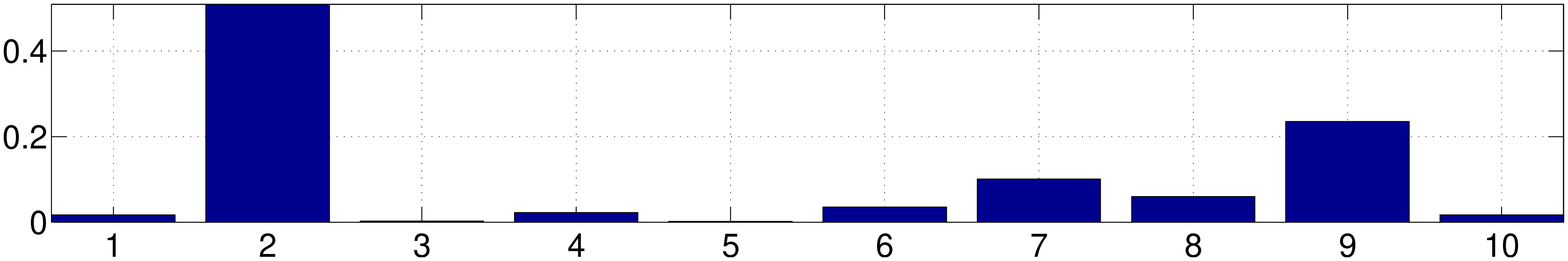}
\vspace*{2mm}
\caption{$\beta_k$ vs $k$}
\label{anec_beta}
\end{center}
\end{subfigure}
\caption{({\it Color Online}): Performance on mixture speech signal with 2 active speakers.}
\label{anech_res}

\end{figure}

As an illustration, Fig.\ref{anech_res} shows the plots of ground truth of who spoke what (Fig. \ref{anec_gnd_trth}),  $JPM\left(k,l\right)$ (Fig. \ref{anec_delta}) and $\underline{\beta}$ (Fig. \ref{anec_beta})  on a sample scalar mixture with two active speakers ($S_2$ and $S_9$) speaking simultaneously. From Fig. \ref{anec_delta}, we see that LVDSK gives higher probability values for the correct speaker speaking the correct word and lower probability values for other combinations. It can also be seen that the $\underline{\beta}$ values (Fig.\ref{anec_beta}) yield higher probability for the active speakers and lower probability for the inactive speakers. Thus the active speakers are accurately identified in the plot of $\underline{\beta}$.

\subsubsection{Performance under different Tasks}
\label{sec:perf:res:tasks}

Table. \ref{tab_tasks} tabulates the percentage recognition accuracies for the two tasks and the overall performance. LVDSK has near perfect performance for recognition of at least one of the two entities (speaker, keyword or speaker-keyword pair). The performance of recognizing both the speakers is higher in the MSpSKW task than the MSpDKW task. This is intuitively reasonable because, one would expect to differentiate speakers easily when both are uttering the same content, versus both speakers uttering different content. 

With just simple mixture models, the recognition of both the keywords, is more than $80\%$. In the case of MSpSKW task, in which the same keyword is uttered by both speakers, although the confusability is higher, when compared to the MSpDKW task, LVDSK performs quite well with more than $90\%$ recognition accuracy for detecting both the keywords. This is because, most of the errors occurring in the MSpDKW task are when a smaller keyword - like ``Hello" or ``Music" is completely embedded within a larger keyword like ``Emergency" or ``Disconnect". Another source of error observed, is the confusion between the words ``Inside" and ``Outside", when the initial part of these words are masked by another keyword. These scenarios never occur in the MSpSKW task as the keywords are roughly of the same length and spectral content. The words, ``Answer" and ``Number" are usually accurately detected, except when their initial part is masked. As expected, the longer distinct keywords, are least confused. 

If the speakers and the keywords are correctly detected, but they are wrongly mapped to each other, then the recognition of both speaker-keyword pairs, will be incorrect. Thus, one can expect the accuracy for detecting both speaker-keyword pairs to be lower than the lesser of the recognition accuracies for detecting both speakers and both keywords. The overall performance is computed on combining the MSpDKW and the MSpSKW tasks. Since, the MSpDKW task has more number of utterances, the overall performance of LVDSK framework is closer to MSpDKW than the MSpSKW task.

In general, the speaker recognition accuracy is higher than keyword recognition accuracy owing to the use of mixture models (and the assumption of i.i.d. features, thereof). With more sophisticated keyword models (Long Short Term Memory (LSTM) recurrent neural networks \cite{Wollmer13}), it may be possible to achieve higher accuracy for recognizing both the keywords. More investigations regarding the suitable choice of speaker-specific keyword models are warranted, and this is beyond the scope of this paper. Importantly, any such models (LSTM networks) can be used in the proposed framework.

\subsubsection{Performance with Prior Knowledge}
\label{sec:perf:res:prior}
Let the scenario in which, the prior knowledge of active speakers is used in the LVSDK framework (as shown in Sec. \ref{sec:prior}), be referred to as Oracle speaker identities (Oracle-SpID). Similarly, the scenario in which the prior knowledge of keywords is incorporated into LVDSK is referred to as Oracle keyword identities (Oracle-KWID). Table III tabulates the performance of LVDSK under the Oracle-SPID and Oracle-KWID scenarios. Clearly, using the prior knowledge of the active speakers boosts the recognition accuracy of keywords and hence the speaker-keyword pairs. Similarly, using the prior knowledge of keywords, boosts the recognition accuracy of active speakers. It is observed that the prior knowledge of keywords is more informative to the LVDSK framework than the prior knowledge of the speakers. This can be attributed to the mixture models being better at modeling speakers than keywords. 

\section{Conclusions}
\label{conclu}

In this paper we have proposed a LV framework to address the problem of detecting multiple speakers and their keywords in a multi-speaker mixture signal. The proposed framework is generic enough to incorporate any kind of speaker specific keyword models. Analysis of LVDSK with GMMs and tMMs as speaker-specific-keyword models showed the superior performance of tMMs over GMMs. LVDSK also offers an elegant way to incorporate prior knowledge about the speaker and/or the keywords.  With accurate prior knowledge of the keywords a significant improvement in the recognition of speaker-keyword pairs is achieved.


\bibliographystyle{IEEEbib}
\bibliography{Keywrd_spid}

\end{document}